\documentclass[twocolumn,floatfix,showpacs]{revtex4}

\usepackage{amsfonts}
\usepackage{amsmath}
\usepackage{amssymb}
\usepackage{mathrsfs}
\usepackage[english,activeacute]{babel}
\usepackage[latin1]{inputenc}
\usepackage{graphicx}
\usepackage{subfigure}
\usepackage{epsfig}
\usepackage{float}

 \usepackage{color}

\begin{document}

\title{Extension of a factorization method of nonlinear second order ODE's with variable coefficients}

\author{H.C. Rosu}\email{hcr@ipicyt.edu.mx}
\affiliation{IPICyT, Instituto Potosino de Investigacion Cientifica y Tecnologica,\\
Camino a la presa San Jos\'e 2055, Col. Lomas 4a Secci\'on, 78216 San Luis Potos\'{\i}, S.L.P., Mexico}%
\author{O. Cornejo-P\'erez}\email{octavio.cornejo@uaq.mx}
\affiliation{Facultad de Ingenier\'{\i}a, Universidad Aut\'onoma de Quer\'etaro,\\
Centro Universitario Cerro de las Campanas, 76010 Santiago de Quer\'etaro, Mexico}
\author{M. P\'erez-Maldonado}\email{maximino.perez@ipicyt.edu.mx}
\affiliation{IPICyT, Instituto Potosino de Investigacion Cientifica y Tecnologica,\\
Camino a la presa San Jos\'e 2055, Col. Lomas 4a Secci\'on, 78216 San Luis Potos\'{\i}, S.L.P., Mexico
}%
\author{J.A. Belinch\'on}\email{abelcal@ciccp.es}
\affiliation{Departamento de F\'{\i}sica, Facultad de Ciencias Naturales,\\ Universidad de Atacama, Copayapu 485
Copiap\'o, Chile}

\pacs{02.30.Hq; 11.30.Pb\hfill arXiv:1612.01938v3}

\begin{abstract}
\noindent The factorization of nonlinear second-order differential equations proposed by Rosu and Cornejo-P\'erez in 2005 is extended to equations
containing quadratic and cubic forms in the first derivative. A few illustrative examples encountered in physics are provided.\\

\noindent {\bf Keywords}: nonlinear second order equation; factorization; powers of first derivative

%
%
\end{abstract}

\centerline{\hspace{9.5cm} Rev. Mex. F\'{\i}s. 63 (2017) 218-222}
\centerline{}
\centerline{}
\centerline{}
\centerline{}

\maketitle

\section{Introduction}

Finding exact solutions of nonlinear differential equations
has long been an active field of research because of the insight
they offer in the understanding of many processes in
physics, biology, chemistry, and other scientific areas.
Among the methods developed to find analytical solutions of nonlinear
ordinary differential equations (ODEs) and nonlinear partial differential equations
(PDEs) we enumerate the truncation procedure in Painlev\'{e}
analysis \cite{weiss}, the Hirota bilinear method \cite{hirota},
the tanh function method \cite{parkes,fan}, the Jacobi elliptic
function method \cite{fu}, and the Prelle-Singer method
\cite{prelle,chandra2}.

The factorization method, which in mathematics has roots that go to Euler and Cauchy, is a well-known
technique used to find exact solutions of linear second order ODEs
in an algebraic manner. In physics, it has attracted much interest as an elegant way of solving fundamental eigenvalue problems
in quantum mechanics, and later due primarily to its natural association with supersymmetric quantum mechanics \cite{schroed,IH,Mi,Fe,suk85,BO,Dong}.
The latter approach has been extended to some types of nonlinear ODEs \cite{AI12}, and to more dimensions \cite{abi84,i04,i06,i09} as well.
In recent times, the factorization technique has been applied to find exact solutions
of many nonlinear ODEs \cite{berk}, and to nonlinear PDEs, mainly in the context of traveling waves
\cite{bookb,rosu1,cornejo1,vall,estevez1,estevez2,wang,fahmy1,MR}.
The factorization technique was further extended to a class
of coupled Li\'enard equations, which also included a coupled version of the modified Emden equation,
by Hazra {\em et al} \cite{hazra}. Their algorithm can be generalized to higher order scalar and
coupled ODEs, but one has to pay the price of increased algebraic complexity.
In addition, Tiwari {\em et al} \cite{tiwari} factorized even more complicated
quadratic and mixed Li\'enard-type nonlinear systems, among which the coupled Mathews-Lakshmanan nonlinear oscillators.

In this paper, we generalize the factorization technique that we introduced previously \cite{rosu1,cornejo1} for nonlinear equations
with a monomial function in the first derivative, {\em i.e.}, with a damping term which can be also nonlinear,
to nonlinear equations with polynomial functions of second and third degree in the first derivative.
In the following section, we review the factorization in the monomial case. Next, we present the factorization of nonlinear
equations with polynomial function of second degree in the first derivative and illustrate it with a couple of examples.
The last section is devoted to the factorization of nonlinear equations with polynomial function of third degree in the
first derivative. We end up the paper with the conclusion section.

\section{Factorization of nonlinear equations with a monomial of first degree in the first derivative}

Nonlinear equations of the type
\begin{equation}\label{n1}
y_{ss}+f(y,s) y_s +F(y,s)=0~,
\end{equation}
where the subscript $s$ denotes the derivative with respect to $s$ and $F(y,s)$ and $f(y,s)$ are arbitrary functions of $y(s)$ and $s$, 
necessarily a polynomial function
can be factorized as follows \cite{est11}:
\begin{equation}
[D_s-\phi_{2}({y,s})][D_s-\phi_{1}(y,s)]y(s)=0~,\label{n2}
\end{equation}
where $D_{s}=\frac{d}{ds}$.
Expanding (\ref{n2}), one can use the following grouping of terms \cite{rosu1,cornejo1}:
\begin{equation}\label{n4}
D^{2}_{s} y -\left(\phi_{1}+\phi_{2}+
\frac{d\phi_{1}}{dy} y \right)D_s y+\left(\phi_{1}\phi_{2}-\partial \phi_1/\partial s\right)y=0~,
\end{equation}
and comparing Eq.~(\ref{n1}) with Eq.~(\ref{n4}), we get the conditions
\begin{eqnarray}
&&\phi_1+\phi_{2}+ \frac{\partial \phi_1}{ \partial y}y= -f~,\label{n5}\\
&&\phi_{1}\, \phi_{2}-\frac{\partial \phi_1}{\partial s}=\frac{F(y,s)}{y}~.\label{n6}
\end{eqnarray}
Any factorization like (\ref{n2}) of a scalar equation of the form given in Eq.~(\ref{n1}) allows us to find a compatible first order nonlinear
differential equation,
\begin{equation}\label{n7}
[D_s-\phi_{1}(y,s)]y\equiv D_s y-\phi_{1}(y,s)y=0~,
\end{equation}
whose solution provides a particular solution of (\ref{n1}).
In other words, if we are able to find a couple of functions $\phi_{1}(y,s)$ and $\phi_{2}(y,s)$ such that
they factorize Eq.~(\ref{n1}) in the form (\ref{n2}), solving Eq.~(\ref{n7}) allows to get particular solutions
of (\ref{n1}). The advantage of this factorization has been shown in the important particular case when there is
no explicit dependence on $s$, {\em i.e.}, for equations
\begin{equation}\label{n1b}
y_{ss}+f(y) y_s +F(y)=0,
\end{equation}
for which the factorization conditions are
\begin{eqnarray}
&&\phi_1+\phi_{2}+ \frac{d\phi_{1}}{dy}y= -f~,\label{n5b}\\
&&\phi_{1}\, \phi_{2}=\frac{F(y)}{y}~,\label{n6b}
\end{eqnarray}
when the two unknown functions $\phi_{1}(y)$ and $\phi_{2}(y)$ can be found easily by factoring $F(y)$ when
it is a polynomial or written as a product of two functions. This property of the nonlinear factorization has been
successfully used when it has been introduced a decade ago and contributed to its popularity \cite{GS}.
An illustration of this technique in the case of
the cubic Ginzburg-Landau equation can be found in \cite{kink2012}.
Notice that interchanging the factoring functions turns
(\ref{n5b}) and (\ref{n6b}) into
\begin{eqnarray}
&&\phi_1+\phi_{2}+ \frac{d\phi_{2}}{dy}y= -\tilde{f}~,\label{n5bb}\\
&&\phi_{1}\, \phi_{2}=\frac{F(y)}{y}~,\label{n6bb}
\end{eqnarray}
which correspond to equations
\begin{equation}\label{n1bb}
y_{ss}+\tilde{f}(y) y_s +F(y)=0~.
\end{equation}
If $s$ is a traveling variable, this suggests kinematic relationships between the kink solutions of (\ref{n1b}) and (\ref{n1bb}) evolving under the
different nonlinear dampings $f(y)$ and $\tilde{f}(y)$.

\medskip

Finally, in the case $f=0$ and $F(y,s)=V(s)y$, the factoring functions $\phi$'s depend only on $s$ and
the equations (\ref{n1}) are linear ones
\begin{equation}\label{n1c}
y_{ss}+V(s)y=0~.
\end{equation}
The factorization conditions take the simplified form
\begin{eqnarray}
&&\phi_1+\phi_{2}= 0~,\label{n5c}\\
&&\phi_{1}\, \phi_{2}-\frac{d \phi_1}{d s}=V(s)~.\label{n6c}
\end{eqnarray}
From (\ref{n5c}), one has $\phi_1=-\phi_2=\phi$ which upon substitution in (\ref{n6c}) leads to the well known Riccati
equation $-d\phi/ds-\phi^2=V(s)$ defining the Schr\"odinger potential in quantum mechanics in terms of the factoring function. The interchange of
$\phi_1$ with $\phi_2$ produces the partner Riccati equation $d\phi/ds-\phi^2=\tilde{V}(s)$ of much use in supersymmetric quantum mechanics
\cite{Bag,Coop}.

\bigskip

\section{Factorization of nonlinear equations with polynomial function of second degree in the first derivative}

Let us consider the following nonlinear second order ODE with variable coefficients
\begin{equation}
y_{ss} +f(y,s)y_{s}^{\,2} + g(y,s) y_s +F(y,s) =0~. \label{eq1a}%
\end{equation}
A factorization of the form
\begin{equation}
\left[  D_{s} +f(y,s)y_s -\phi_{2}(y,s) \right]  \left[  D_{s}-\phi_{1}(y,s)\right]  y=0~, \label{eq2b}%
\end{equation}
is possible if the following constraint equations are satisfied:
\begin{eqnarray}
&& \phi_{1} +\phi_{2} +\left(\frac{\partial\phi_{1}}{\partial y}+f(y,s) \phi_{1}\right)y  =  -g(y,s),\label{eq4b}\\
&&\phi_{1} \phi_{2} -\frac{\partial\phi_{1}}{\partial s} = \frac{F(y,s)}{y}~ \label{eq4c}
\end{eqnarray}
There are also cases when one can work with $\phi_2=0$. In such cases, the constraint equations take the form
\begin{eqnarray}
&& \phi_{1} +\left(\frac{\partial\phi_{1}}{\partial y}+f(y,s) \phi_{1}\right)y = -g(y,s),\label{eq4bb}\\
&& -\frac{\partial\phi_{1}}{\partial s} = \frac{F(y,s)}{y}~. \label{eq4cc}
\label{eq5b}%
\end{eqnarray}

Finally, the degenerate case corresponding to $\phi_1=0$, which also implies $F=0$, leads to the simple constraint
\begin{equation}\label{d1}
\phi_2=-g(y,s)~.
\end{equation}
As an example of a degenerate case, we mention the equation for the radial function of the isotropic metric in general relativity \cite{Buch1}
\begin{equation}\label{B1}
y_{ss}-\frac{3}{y}y_{s}^{\,2}-\frac{1}{s}y_s=0~,
\end{equation}
for which (\ref{d1}) is written as
\begin{equation}\label{B2}
\phi_{2} = \frac{1}{s}~.
\end{equation}
The solution
\begin{equation}\label{B3}
y=\frac{1}{2}\frac{a}{\sqrt{1+bs^2}}~,
\end{equation}
where $a$ and $b$ are integration constants, can be found by elementary means \cite{Buch1}.

\bigskip

The most important application is when no explicit dependence on $s$ occurs in the equation and so neither $F$ nor the $\phi$'s
depend on $s$ when the constraints are similar to (\ref{n5b}) and (\ref{n6b}). If moreover one assumes $\phi_1=\phi_2=\phi$ then the second constraint
equation provides the factorization function as
\begin{equation}\label{C1}
\phi(y)=\sqrt{\frac{F(y)}{y}}~.
\end{equation}
Substituting (\ref{C1}) in the first constraint equation leads to the following expression for the $g$ coefficient
\begin{equation}\label{C3}
g(y)=-\frac{1}{2}\sqrt{\frac{F(y)}{y}}\bigg[3+\left(\frac{F_y}{F}+2f(y)\right)y\bigg]~.
\end{equation}
For given $f(y)$ and $F(y)$, the latter equation gives the coefficient $g(y)$ for which the nonlinear equation can be factorized in the form
\begin{equation}\label{C4}
\left[D_{s} +f(y)y_s - \sqrt{F(y)/y}\right] \left[D_{s}-\sqrt{F(y)/y}\right]  y=0~.%
\end{equation}
There are equations of the latter type which do not present a linear term in the first derivative. This implies $g(y)=0$, {\em i.e.}
\begin{equation}\label{C5}
3+\left(\frac{F_y}{F}+2f(y)\right)y=0~,
\end{equation}
which is separable. The solution
\begin{equation}\label{C6}
F(y)=Cy^{-3}e^{-2\int^yf(u)du}~,
\end{equation}
with $C$ an integration constant, provides the form of $F$ which for given $f$ allows the factorization of the equation.
However, as simple as it may look, the condition (\ref{C6}) is quite restrictive.

In physical applications, differential equations with squares of the first derivative are encountered in highly nonlinear areas, such as cosmology
\cite{mhb} and gravitation theories, {\em e.g.}, Weyl conformal gravity \cite{Deliduman} and $f(R)$ gravity \cite{Bohmer}, but occasionally they show
up in other branches as well. In the following, we will give two examples of factorization of such equations.

\subsection{An equation in Weyl's conformal cosmology}


The following equation
\begin{equation}
y_{ss}-\frac{\alpha}{y}y_{s}^{\,2}+\frac{y^{\sigma}}{x^{2}}=0~,
\label{Peq1b}%
\end{equation}
where $\alpha$ and $\sigma$ are real constants,
arises in intermediate calculations concerning the vacuum
solution of the field equations in Weyl's conformal gravity \cite{poveromo,mannheim}.
Let us try the factorization
\begin{equation}
\left(  D_{s}-\frac{\alpha}{y}y_s\right)  \left(D_{s}-\phi_{1}(y,s)\right)y=0~. \label{Peq2b}%
\end{equation}
Therefore, the following constraint equations should be satisfied
\begin{eqnarray}
&&\frac{\partial\phi_{1}}{\partial s}=-\frac{y^{\sigma-1}}{s^{2}%
}\label{Peq5b}\\
&&\phi_{1}-\frac{\alpha}{y}\phi_{1}y+\frac{\partial\phi_{1}}{\partial y}y=0.
\label{Peq6b}%
\end{eqnarray}
Equation (\ref{Peq6b}) is separable and generates the function $\phi_{1}(y,s)=f(s)y^{\alpha-1}$,
then, from eq.~(\ref{Peq5b}) we obtain
\begin{equation}
\frac{\partial\phi_{1}}{\partial s}=\frac{\partial}{\partial s}(y^{\alpha
-1}f(s))=y^{\alpha-1}f^{\prime}(s)=-\frac{y^{\sigma-1}}{s^{2}} \label{Peq7b}%
\end{equation}
which implies $\alpha=\sigma$, and $f(s)=\frac{1}{s}+c_{1}$, where $c_{1}$ is an arbitrary constant.

Assuming the following \cite{wang}
\begin{equation}
\left(  D_{s}-\phi_{1}(y,s)\right)y=\Omega, \label{Peq8b}%
\end{equation}
then, we get
\begin{equation}
\Omega^{\prime}-\alpha\frac{y^{\prime}}{y}\Omega=0. \label{Peq9b}%
\end{equation}
with solution $\Omega=k_{0}y^{\alpha}$. Therefore, we get the first order
equation
\begin{equation}
y^{\prime}-\left(  \frac{1}{s}+c_{1}\right)  y^{\alpha}=k_{0}y^{\alpha},
\label{Peq10b}%
\end{equation}
which can be rewritten in the form
\begin{equation}
y^{\prime}-\left(\frac{1}{s}+k_{1}\right)  y^{\alpha}=0~, \label{Peq11b}%
\end{equation}
where $k_{1}$ is an integration constant. The general solution of Eq.~(\ref{Peq11b}) is given in the form
\begin{equation}
y=((\alpha-1)\left(-k_{1}s-k_{2}-\ln s)\right)^{\frac{1}{1-\alpha}}.
\label{Peq12b}%
\end{equation}
where $k_{2}$ is an integration constant. For $k_{1}=0$ and $\alpha=5$, we obtain the following particular solutions
\begin{eqnarray}
y_{1,2}  &  =\pm\frac{1}{\sqrt{2}(-k_{2}-\ln(s))^{1/4}}~,\label{Peq13b}\\
y_{3,4}  &  =\pm\frac{i}{\sqrt{2}(-k_{2}-\ln(s))^{1/4}}~. \label{Peq14b}%
\end{eqnarray}

\subsection{Langmuir-type equations}

A particular example of the type (\ref{eq1a}) is the following equation
$$
\frac{d^2y}{ds^2}+\frac{1}{3y}\left(\frac{dy}{ds}\right)^2+\gamma\frac{dy}{ds}
+\frac{1}{3}\frac{y^2-1}{y}=0~
$$
which when $\gamma=4/3$ provides Langmuir's radial $\beta$ function occurring in the formula for the space charge between coaxial cylinders
\cite{Lang23}.
Using (\ref{eq4c}), one can choose
\begin{equation}\label{L1}
\phi_1=-\left(1-\frac{1}{y}\right)~, \qquad \phi_2=-\frac{1}{3}\left(1+\frac{1}{y}\right)~.
\end{equation}
Substituting (\ref{L1}) in (\ref{eq4b}), one obtains $\gamma=5/3$, which shows that the Langmuir case cannot be factored.
If $\gamma=5/3$, we can obtain a particular solution from the first-order differential equation
\begin{equation}\label{Lp}
\left(D_{s}+1-\frac{1}{y}\right)y=0~ \Longrightarrow y_s+y-1=0~,
\end{equation}
which is
\begin{equation}\label{Lp1}
y(s)=Ce^{-s}+1~,
\end{equation}
where $C$ is the integration constant.


\bigskip



\section{Factorization of nonlinear equations with polynomial function of third degree in the first derivative}

It is well known that equations of the type
\begin{equation}
y_{ss} +f(y,s) y_{s}^{\,3} + g(y,s)y_{s}^{\,2}+h(y,s)y_s+F(y,s)=0~, \label{equ1}%
\end{equation}
where the coefficient functions are mappings from two-dimensional disks to the set of real numbers, ${\cal D}^2\rightarrow \mathbb{R}$, define
projective connections \cite{Matveev,Bagderina}.

Such equations allow for the factorization
\begin{equation}
\left[D_{s}+f(y,s) \dot{y}^2 -\phi_2(y,s) \right]  \left[  D_{s}-
\phi_{1}(y,s)\right]  y=0~, \label{equ2}%
\end{equation}
with the compatible first order equation
\begin{equation}
 \left[D_{s}-\phi_{1}(y,s)\right]y\equiv y_s -\phi_1(y,s) y =0~, \label{equ2a}
\end{equation}
under the constraint equations
\begin{eqnarray}
&&f(y,s)\phi_{1}y  =  -g(y,s)\label{equ3}\\
&&\phi_1+\phi_{2} +\frac{\partial \phi_{1}}{\partial y}y  =  -h(y,s) \label{equ4}\\
&&\phi_{1} \phi_{2}-\phi_{1s}y = \frac{F(y,s)}{y}~. \label{equ5}
\end{eqnarray}

On the other hand, for any symmetric affine connection $\Gamma =(\Gamma^i_{jk}(s,y))$,
the so-called projective connection \emph{associated to}~$\Gamma$ \cite{Matveev} which carries all information about unparametrized geodesics of
$\Gamma$ is determined by the equation
\begin{equation} \label{cub-con}
y_{ss}-\Gamma^1_{22}y_{s}^{\,3}+(\Gamma^2_{22}{-}2\Gamma^1_{12})y_{s}^{\,2} -(\Gamma^1_{11}{-}2\Gamma^2_{12})\,y_s+\Gamma^2_{11}=0~.
\end{equation}
Thus, one finds that equations (\ref{cub-con}) can be factored if
\begin{eqnarray}
&&\phi_{1}y =  (\Gamma^2_{22}{-}2\Gamma^1_{12})/\Gamma^1_{22} \label{equc3}\\
&&\phi_1+\phi_{2} +\frac{\partial \phi_{1}}{\partial y}y  = \Gamma^1_{11}{-}2\Gamma^2_{12} \label{equc4}\\
&&\phi_{1} \phi_{2}-\phi_{1s}y = \frac{\Gamma^2_{11}}{y}~. \label{equc5}
\end{eqnarray}

We do not present any particular case. Rather we notice that for given $\Gamma$'s,
(\ref{equc3}) provides $\phi_1$. Then, substituting in (\ref{equc4}), we get $\phi_2$, but in the end (\ref{equc5}) should be still satisfied.
This looks complicated and makes the success of the method less probable.

\section{Conclusion}

\medskip

In summary, we have discussed here a simple factorization method of complicated nonlinear second-order differential equations containing
quadratic and cubic polynomial forms in the first derivative, and we have presented some examples.
Only those equations with the coefficients satisfying certain constraints involving
the factoring functions can be factorized. By doing this, one can seek solutions of simpler first
order nonlinear differential equations, corresponding to the first factorization bracket from the right.
This works fine when there is only a linear term in the first derivative. When the powers of the first derivatives
are more than one, the constraint conditions on the factoring functions become more complicated, and the factorization method
is less appropriate. In general, the factorization method can still work when the coefficients of the nonlinear equation do not
depend explicitly on the independent variable, because the constraint equations are less restrictive in these cases.

\bigskip
\bigskip

\newpage


{\bf Acknowledgements}

\medskip

The authors wish to thank Dr. J. Poveromo for informing them on the nonlinear equation occurring in Weyl's conformal gravity model.
M. P\'erez-Maldonado thanks CONACyT for a doctoral fellowship.


\bigskip
\bigskip


%



\begin{thebibliography} {99}

\bibitem{weiss} J. Weiss,
\textit{J. Math. Phys.} {\bf 24} (1983) 1405.

\bibitem{hirota} R. Hirota,
\textit{Phys. Rev. Lett.} {\bf 27} (1971) 1192.

\bibitem{parkes} E.J. Parkes and B.R. Duffy,
\textit{Comput. Phys. Commun.} {\bf 98} (1996) 288.

\bibitem{fan} E.G. Fan,
\textit{Phys. Lett. A} {\bf 277} (2000) 212.

\bibitem{fu} Z.T. Fu, S.K. Liu, S.D. Liu, and Q. Zhao,
\textit{Phys. Lett. A} {\bf 290} (2001) 72.

\bibitem{prelle} M. Prelle and M. Singer,
\textit{Trans. Am. Math. Soc.} {\bf 279} (1983) 215.

\bibitem{chandra2} V.K. Chandrasekar, S.N. Pandey, M. Senthilvelan, and M. Lakshmanan,
\textit{J. Math. Phys.} {\bf 47} (2006) 023508.

\bibitem{schroed} E. Schr\"{o}dinger,
\textit{Proc. R. Ir. Acad. A} {\bf 47} (1941-1942) 53.

\bibitem{IH} L. Infeld and T.E. Hull,
\textit{Rev. Mod. Phys.} {\bf 23} (1951) 21.

\bibitem{Mi} B. Mielnik,
\textit{J. Math. Phys.} {\bf 25} (1984) 3387.

\bibitem{Fe} D.J. Fern\'andez C.,
\textit{Lett. Math. Phys.} {\bf 8} (1984) 337.

\bibitem{suk85} C.V. Sukumar,
\textit{J. Phys. A: Math. Gen.} {\bf 18} (1985) L57.

\bibitem{BO} B. Mielnik and O. Rosas-Ortiz,
\textit{J. Phys. A} {\bf 37} (2004) 10007.

\bibitem{Dong} S.-H. Dong, {\em ``Factorization Method in Quantum Mechanics"}, Springer, (2007).

\bibitem{AI12} A.A. Andrianov and M.V. Ioffe,
\textit{J. Phys. A: Math. Theor.} {\bf 45} (2012) 503001.

\bibitem{abi84} A.A. Andrianov, N.V. Borisov, and M.V. Ioffe,
\textit{JETP Lett.} {\bf 39} (1984) 93.

\bibitem{i04} M.V. Ioffe,
\textit{J. Phys. A: Math. Gen.} {\bf 37} (2004) 10363.

\bibitem{i06} M.V. Ioffe, J. Negro, L. M. Nieto, and D. N. Nishnianidze,
\textit{J. Phys. A: Math. Gen.} {\bf 39} (2006) 9297.

\bibitem{i09} F. Cannata, M.V. Ioffe, and D. N. Nishnianidze,
\textit{J. Math. Phys.} {\bf 50}, (2009) 052105.

\bibitem{berk} L.M. Berkovich,
\textit{Appl. Anal. Discrete Math.} {\bf 1} (2007) 122.


\bibitem{bookb} L.M. Berkovich, {\em ``Factorizations and Transformations of Differential Equations"},
Regular and Chaotic Dynamics Editorial Center, (in Russian), (2002).


\bibitem{rosu1} H.C. Rosu and O. Cornejo-P\'erez,
\textit{Phys. Rev. E} {\bf 71} (2005) 046607.

\bibitem{cornejo1} O. Cornejo-P\'erez and H.C. Rosu,
\textit{Prog. Theor. Phys.} {\bf 114} (2005) 533.

\bibitem{vall} O. Cornejo-P\'erez, J. Negro, L.M. Nieto, and H.C. Rosu,
\textit{ Found. Phys.} {\bf 36} (2006) 1587.

\bibitem{estevez1} P.G. Est\'evez, \c S. Kuru, J. Negro, and L.M. Nieto,
\textit{J. Phys. A: Math. Gen.} {\bf 39} (2006) 11441.

\bibitem{estevez2} P.G. Est\'evez, \c S. Kuru, J. Negro, and L.M. Nieto,
\textit{J. Phys. A: Math. Theor.} {\bf 40} (2007) 9819.

\bibitem{wang} D.S. Wang and H. Li,
\textit{J. Math. Anal. Appl.} {\bf 343} (2008) 273.

\bibitem{fahmy1} E.S. Fahmy,
\textit{Chaos, Solitons and Fractals} {\bf 38} (2008) 1209.

\bibitem{MR} S.C. Mancas and H.C. Rosu,
\textit{Phys. Lett. A} {\bf 377} (2013) 1434.

\bibitem{hazra} T. Hazra, V. K. Chandrasekar, R. Gladwin Pradeep, and M. Lakshmanan,
\textit{J. Math. Phys.} {\bf 53} (2011) 023511.

\bibitem{tiwari} A.K. Tiwari, S.N. Pandey, V.K. Chandrasekar, and M. Lakshmanan,
\textit{Appl. Math. Comp.} {\bf 252} (2015) 457.

\bibitem{est11} P.G. Est\'evez, \c S. Kuru, J. Negro, and L.M. Nieto,
\textit{Int. J. Theor. Phys.} {\bf 50} (2011) 2046.

\bibitem{GS} G.W. Griffiths and W.E. Schiesser, {\em ``Traveling Wave Analysis of Partial Differential Equations,
Numerical and Analytical Methods with MATLAB and MAPLE"}, Academic Press, (2012).

\bibitem{kink2012} H.C. Rosu, O. Cornejo-P\'erez, P. Ojeda-May,
\textit{Phys. Rev. E} {\bf 85} (2012) 037102.

\bibitem{Bag} B. Bagchi, {\em ``Supersymmetry in Quantum and Classical Mechanics"}, Chapman and Hall/CRC, (2001).

\bibitem{Coop} F. Cooper, A. Khare, U. Sukhatme, {\em ``Supersymmetry in Quantum Mechanics"}, World Scientific, (2001).

\bibitem{Buch1} H.A. Buchdahl,
\textit{Astrophys. J.} {\bf 140}, (1964) 1512.

\bibitem{mhb} M.K. Mak, T. Harko, and J.A. Belinch\'on,
\textit{Int. J. Mod. Phys. D} {\bf 11} (2002) 1265.

\bibitem{Deliduman} C. Deliduman, O. Kasikci, and C. Yapiskan,
Flat galactic rotation curves from geometry in Weyl gravity,
arXiv:1511.07731.

\bibitem{Bohmer} C.G. B\"ohmer, T. Harko, and F. S. N. Lobo,
\textit{Astropart. Phys.} {\bf 29} (2008) 386.

\bibitem{poveromo}  J. Poveromo, private communication.

\bibitem{mannheim}  P.D. Mannheim and D. Kazanas,
\textit{Astrophys. J.} {\bf 342} (1989) 635.

\bibitem{Lang23} I. Langmuir and K.B. Blodgett,
\textit{Phys. Rev.} {\bf 22} (1923) 347.

\bibitem{Matveev} V.S. Matveev,
\textit{Math. Ann.} {\bf 352} (2012) 865.

\bibitem{Bagderina} Y.Y. Bagderina,
\textit{J. Appl. Ind. Math.} {\bf 10} (2016) 37.



\end{thebibliography}
\end{document}